\documentclass{emulateapj}

\pdfoutput=1
\usepackage{epsf}
\usepackage{epsfig}
\usepackage{natbib}
\DeclareGraphicsExtensions{.eps,.jpg,.pdf,.png}

\usepackage{color}
\def\red#1 {\textcolor{red}{#1}\ }

\newcommand{\chandra}{{\it Chandra}}

\newcommand{\Msun}{M$_\odot$}
\newcommand{\bq}{\begin{equation}}
\newcommand{\eq}{\end{equation}}

\defcitealias{Allen02}{A02}

\shorttitle{CARMA Measurements of the Sunyaev-Zel'dovich Effect in RX~J1347.5$-$1145}
\shortauthors{Plagge et al.}

\begin{document}

\title{CARMA Measurements of the Sunyaev-Zel'dovich Effect in RX~J1347.5$-$1145}

%%
%% Authors
%%
\author{
Thomas~J.~Plagge,$\!$\altaffilmark{1}
Daniel~P.~Marrone,$\!$\altaffilmark{1,2}
Zubair~Abdulla,$\!$\altaffilmark{1}
Massimiliano~Bonamente,$\!$\altaffilmark{3,4}
John~E.~Carlstrom,$\!$\altaffilmark{1,5,6}
Megan~Gralla,$\!$\altaffilmark{1,7}
Christopher~H.~Greer,$\!$\altaffilmark{1}
Marshall~Joy,$\!$\altaffilmark{4}
James~W.~Lamb,$\!$\altaffilmark{8}
Erik~M.~Leitch,$\!$\altaffilmark{1}
Adam~Mantz,$\!$\altaffilmark{1}
Stephen~Muchovej,$\!$\altaffilmark{8}
and David~Woody$\!$\altaffilmark{8}
}

\altaffiltext{1}{Kavli Institute for Cosmological Physics, Department
  of Astronomy and Astrophysics, University of Chicago, Chicago, IL 60637, USA}
\altaffiltext{2}{Steward Observatory, University of Arizona, 933 North Cherry Avenue, Tucson, AZ 85721, USA}
\altaffiltext{3}{Department of Physics, University of Alabama, Huntsville, AL 35899, USA}
\altaffiltext{4}{Space Science-VP62, NASA Marshall Space Flight Center, Huntsville, AL 35812, USA}
\altaffiltext{5}{Enrico Fermi Institute, University of Chicago, Chicago, IL 60637, USA}
\altaffiltext{6}{Department of Physics, University of Chicago, Chicago, IL 60637, USA}
\altaffiltext{7}{Department of Physics and Astronomy, Johns Hopkins University, Baltimore, MD 21218, USA}
\altaffiltext{8}{Owens Valley Radio Observatory, California Institute of Technology, Big Pine, CA 93513, USA}

%%
%% Abstract
%%
\begin{abstract}

We demonstrate the Sunyaev-Zel'dovich (SZ) effect imaging capabilities of the
Combined Array for Research in Millimeter-wave Astronomy (CARMA) by
presenting an SZ map of the galaxy cluster RX~J1347.5$-$1145.  By combining
data from multiple CARMA bands and configurations, we are able to 
capture the structure of this cluster over a wide range of angular 
scales, from its bulk properties to its core morphology.  
We find that roughly 9\% of this cluster's thermal energy is associated 
with sub-arcminute-scale structure imparted by a merger, 
illustrating the value of high-resolution SZ measurements for
pursuing cluster astrophysics and for understanding the scatter in 
SZ scaling relations.
We also find that the cluster's SZ signal is lower in amplitude 
than suggested by a spherically-symmetric model derived from X-ray 
data, consistent with compression
along the line of sight relative to the plane of the sky.
Finally, we discuss the impact of upgrades currently in progress 
that will further enhance CARMA's power as an SZ imaging instrument.

\end{abstract}


\section{Introduction}
\label{sec:intro}

Galaxy clusters are the largest gravitationally bound systems
in the universe, and have taken nearly a Hubble time to form.
They therefore have the potential to act as powerful probes of 
cosmology if systematic errors can be controlled.  Precision 
cluster cosmology will require a deep understanding of cluster
astrophysics, particularly as it relates to the hot gas of the
intracluster medium (ICM). The most detailed studies of the 
ICM have thus far been performed
by X-ray telescopes, which are sensitive to the bremsstrahlung
emission from the $10^7$-$10^8$~K gas.  The Sunyaev-Zel'dovich 
(SZ) effect is a complementary probe of the ICM.
The amplitude of the SZ signal depends on the line-of-sight 
integral of $n_{\mathrm{e}} T$, while the X-ray surface brightness depends on $n_{\mathrm{e}}^2$,
so sensitive SZ measurements can access tenuous gas outside the cluster core and 
directly measure pressure disturbances.
Features found commonly in the outer regions of clusters, 
such as shocked gas from mergers, may therefore be easier 
to detect using the SZ effect than using X-rays.  Moreover, 
the combination of X-ray and SZ data can be 
used to obtain a more complete picture of the ICM thermodynamics. 

To take advantage of these opportunities, 
advances in SZ imaging capabilities are needed.
Measurements of the SZ effect have become routine over the 
last decade, but the full potential of the SZ effect as 
a probe of cluster physics remains largely unexploited due to 
technical challenges: the combination of high sensitivity 
and large angular dynamic range required for detailed SZ imaging has 
proven difficult to achieve with existing instruments.
As a result, the use of the SZ effect has been limited primarily to 
studies where resolved imaging is unnecessary.  

The small number of higher-resolution SZ images obtained to date 
have served to demonstrate the utility of the technique.  However,
single-dish measurements such as those by \citet{Korngut11} can suffer
from radio point source contamination and have been limited 
to scales $<45^{\prime\prime}$ by the necessity of 
filtering out modes contaminated by atmospheric noise.
Multi-dish SZ measurements by arrays such as ATCA \citep[e.g.,][]{Malu10} 
can constrain and remove point sources using the inherent
spatial filtering ability of interferometers, but most
millimeter-wave interferometers lack
sensitivity at arcminute angular scales where the 
SZ cluster signal is largest.

CARMA is a heterogeneous 
interferometric array consisting of 23 antennas with 
diameters of 3.5, 6.1, and 10.4~m operating at 1~cm,
3~mm, and 1~mm.  This particular combination of antennas and
bands makes CARMA a uniquely powerful SZ instrument: its 
3.5~m antennas can be placed in a compact configuration 
sensitive to arcminute-scale emission, and its 6.1
and 10.4~m dishes can be used to obtain the sensitivity
necessary to resolve smaller angular scale SZ 
features.  In this work, we make use of CARMA data from three array
configurations and two bands to obtain an SZ image of the galaxy cluster
RX~J1347.5$-$1145.  These data represent the highest-fidelity
picture of a galaxy cluster ever obtained using the SZ effect.

This paper is organized as follows: 
Section~\ref{sec:rxj1347} provides background on RX~J1347.5$-$1145, 
Section~\ref{sec:obs} describes the observations and data 
reduction, and Section~\ref{sec:decon} discusses the
modeling and deconvolution method.  We present our
results and compare them to previous measurements 
in Section~\ref{sec:results}, and review the conclusions 
and discuss prospects for future work
in Section~\ref{sec:conclusions}. 

\section{RX~J1347.5$-$1145}
\label{sec:rxj1347}

The target for these observations is the cluster RX~J1347.5$-$1145, 
an object that has been characterized extensively using a variety of 
techniques.  First discovered by the ROSAT all-sky survey \citep{ROSAT},
RX~J1347.5$-$1145 is the most luminous cluster known in the X-ray sky, 
and has been measured by several X-ray instruments including 
\emph{Chandra} \citep{Allen02} and XMM-\emph{Newton} \citep{Gitti04}.  
Optical observations have revealed the presence of two cD galaxies,
one coincident with the X-ray emission peak and one directly to the east.
The system has also been found to host a radio mini-halo \citep{Gitti07}. 
Its gravitational potential has been probed using both
strong and weak gravitational lensing \citep[e.g.,][]{Miranda08,Bradac08}.
These multi-wavelength observations indicate that RX~J1347.5$-$1145 is
a massive ($>10^{15}$~\Msun) cluster at redshift $z=0.4510$ which 
has recently undergone a merger with a smaller object.
The cluster's SZ signal has been measured using
single-dish \citep{Komatsu00,Kitayama04,Korngut11} and 
interferometric \citep{Carlstrom02,Bonamente08,Bonamente12} imaging instruments,
and its spectrum near the thermal SZ null has also been characterized
\citep{Zemcov12}.
The higher angular resolution measurements have revealed
a compact region of very hot ($\sim 20$~keV) gas to the
southeast of the X-ray emission peak, while the low-resolution
data indicate a smaller arcminute-scale SZ signal than
suggested by a spherical fit to the X-ray data.

The existence of a central radio-bright AGN, along with the 
limited angular dynamic range of most SZ instruments, has 
complicated efforts to bring SZ data to bear on understanding
this system.  The CARMA data we present help to overcome both
limitations.

\section{Observations and Data Reduction}
\label{sec:obs}

RX~J1347.5$-$1145 was observed with three different sets of CARMA antennas at two wavelengths:
an 8-element sub-array consisting of 3.5~m antennas at 1~cm (``CARMA-8''), a 15-element
sub-array consisting of 6.1~m and 10.4~m antennas at 3~mm (``CARMA-15''), and the full
23-element array at 3~mm (``CARMA-23'').  

The CARMA-8 data were obtained in August 2009 and totaled 25.7~hours of 
unflagged, on-source time.  The center frequency was 31~GHz with a bandwidth
of 8~GHz, and the target R.A.\ and decl.\ were 13:47:30.7 and \mbox{-11:45:08.6} in
J2000 coordinates.  The array was configured with six elements in a
compact array sensitive to arcminute-scale SZ signals, and two
outlying elements providing simultaneous discrimination for compact radio
source emission. The compact array and longer baselines sample
$uv$-spacings of 350$-$1300$~\lambda$ and 2$-$7.5~k$\lambda$,
respectively.  The data were reduced using the 
Sunyaev-Zel'dovich Array (SZA) pipeline described in \citet{Muchovej07}. 

The CARMA-15 data were obtained in July 2010 and totaled 
7.4 hours of unflagged, on-source time.  The center frequency
was 90~GHz with a bandwidth of 8~GHz, and the target R.A.\ and decl.\ were
13:47:32.0 and \mbox{-11:45:42.0} in J2000 coordinates.  The CARMA-15 array was pointed
slightly southeast of the CARMA-8 phase center, directly toward the 
region of hot gas.  The antennas were in the E configuration,
the most compact standard positions for the 6.1 and 10.4~m antennas.
Data reduction was performed using MIRIAD \citep{MIRIAD}.

We obtained CARMA-23 data as part of a commissioning run in February 2011.
All 23 antennas were operated at a center frequency of 86~GHz and were attached to the CARMA spectral line 
correlator, which operated at a reduced bandwidth---4~GHz for the double-sideband receivers
on the 10.4 and 6.1~m antennas, and 2~GHz for the single-sideband receivers
on the 3.5~m antennas.  A total of 8.6~hours of unflagged on-source data were obtained;
a relatively high fraction of the 3.5~m data were flagged due to hardware
issues in the commissioning run that have since been corrected.  The target R.A.\ and decl.\ were
the same as for the CARMA-8 data, and the array configuration was approximately the
combined CARMA-8 and CARMA-15 configurations.  As with the CARMA-15 data,
we reduced the CARMA-23 data using MIRIAD.

For the CARMA-15 and CARMA-23 arrays,
we treat each baseline type (10.4m$\times$10.4m, 10.4m$\times$6.1m, etc.) 
separately to properly account for the differing primary beams.  We therefore
have ten data sets: one for CARMA-8, three for CAMRA-15, and six for CARMA-23.
We apply a cutoff in $uv$ radius for each data set, using the data beyond the 
cutoff only to constrain the point source emission.  The cutoff is
chosen to exclude portions of the $uv$ plane which are poorly sampled for a 
given baseline type.

\begin{figure*}
\begin{center}
\includegraphics*[height=4in]{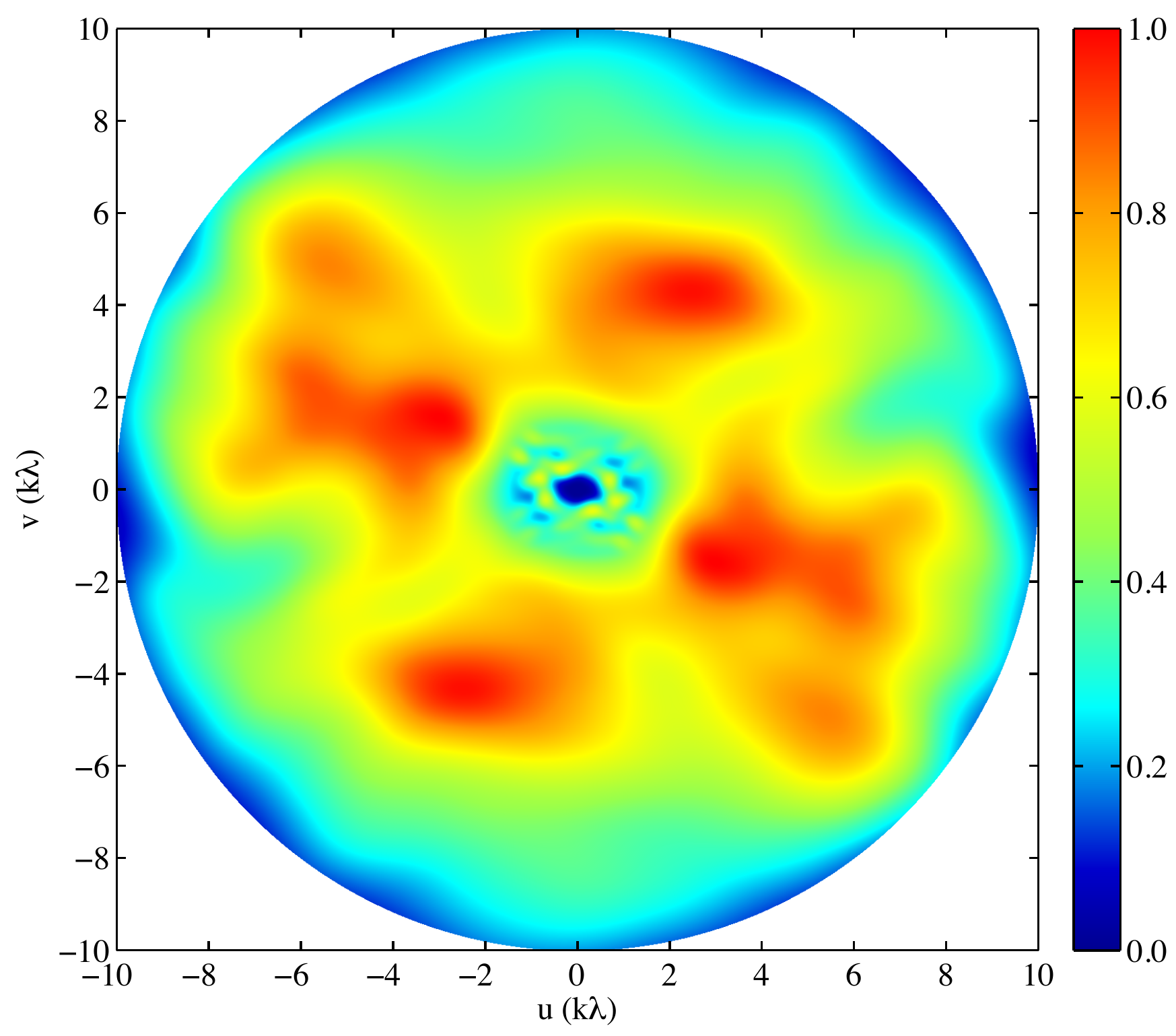}
\caption{Normalized data weight distribution in the $uv$ plane for the union of data sets described in Table~\ref{tab:obs}.
Weights are calculated from the inverse variance of each visibility, scaled by the SZ intensity spectrum. 
The $uv$-plane extent of each visibility weight is determined from
the cross-correlation of the illumination patterns of the corresponding antennas,
providing a more complete view of the $uv$ sampling in the heterogeneous array.  The weights are
well-matched to the cluster signal---which is largest at small $uv$ radius---except for a relatively
under-sampled region around $\sim 2$~k$\lambda$.  This region of the $uv$ plane
will be well-measured by the 23-element CARMA array at $1$~cm, which is currently under development.
\label{fig:uv}}
\end{center}
\end{figure*}

\begin{figure*}
\begin{center}
\includegraphics[height=4in]{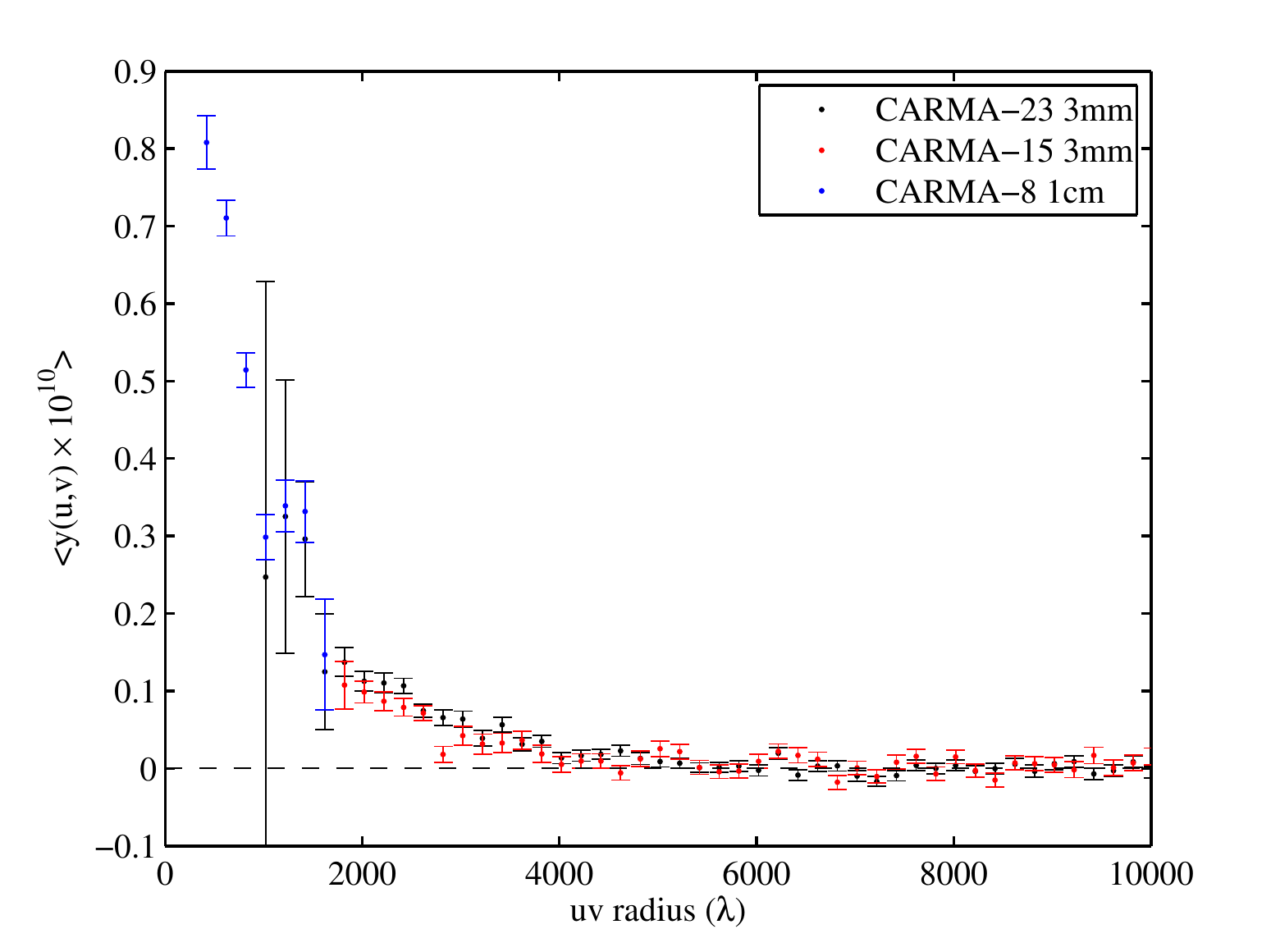}
\end{center}
\caption{Radially binned visibilities from CARMA-8 at 1~cm, CARMA-15 at 3~mm,
and CARMA-23 at 3~mm.  The CARMA-8 data measure the cluster signal at large
angular scales, while the other sub-arrays measure the smaller substructure
of the cluster.}
\label{fig:radbin}
\end{figure*}

The outputs of the data reduction pipelines consist of
flagged, calibrated visibilities $V(u,v)$.  All absolute flux
calibration is performed using the \citet{Rudy87} Mars model,
and is accurate to 5\%.  To combine data from different
bands, we define $y(u,v)$, the Fourier-space counterpart 
to the Compton $y$ parameter \citep{Carlstrom02}:
\begin{equation}
\label{eq:yuv}
y(u,v) \equiv \frac{V_{\nu}(u,v)}{g(\nu,\langle T_{\mathrm{e}} \rangle) \, I_0}
\end{equation}
where $\langle T_{\mathrm{e}} \rangle = 12$~keV is the mean ICM electron temperature of the 
cluster\footnote{We make this approximation due to our 
incomplete of knowledge of the true $T_{\mathrm{e}}(x,y,z)$.  The 
relative value of $g(\nu,T_{\mathrm{e}})$ between 1~cm and 3~mm is nearly independent
of $T_{\mathrm{e}}$, varying by just 2\% between 5 and 15~keV.}.
Our ten $y(u,v)$ data sets are summarized in Table~\ref{tab:obs}.

The data weight distribution in the $uv$ plane for the combined data 
sets is shown in Figure~\ref{fig:uv}.  To illustrate the
contributions of each of the three sub-arrays, we show
the measured $y(u,v)$ binned in $uv$ radius in
Figure~\ref{fig:radbin}.  CARMA-8 at 1~cm constrains 
the cluster at small $uv$ radius (large angular scale)
where the signal is largest, while CARMA-15 and CARMA-23 at
3~mm are sensitive to the large $uv$ radius (small angular scale) 
substructure of the cluster. 

\section{Modeling and Deconvolution}
\label{sec:decon}

We wish to combine the information in all ten sets of visibility data to
form a single image of the cluster.  The first step in this process
is to remove the radio point source emission, which is 
accomplished by fitting a point source model to the 
visibilities above the $uv$ cutoff.
Since radio source fluxes often vary considerably with time, 
identical fluxes should not be expected in the CARMA-15 and CARMA-23 despite their
similar central frequencies.  We therefore demand consistency within each individual 
sub-array, but allow the source flux to vary between sub-arrays.
We find a best-fit centroid consistent with a source in the 
NVSS catalog \citep{NVSS} at an
R.A.\ and decl.\ of 13:47:30.7 and \mbox{-11:45:08.6}.  The flux of this source was
found to be 8.9$\pm$0.5~mJy in the 1~cm CARMA-8 data, 4.2$\pm$0.2~mJy in the 3~mm CARMA-23 data,
and 4.0$\pm$0.2~mJy in the 3~mm CARMA-15 data.  We subtract the best-fit point source
models from the visibility data sets before proceeding.

\begin{figure*}
\begin{center}
\includegraphics[height=5in]{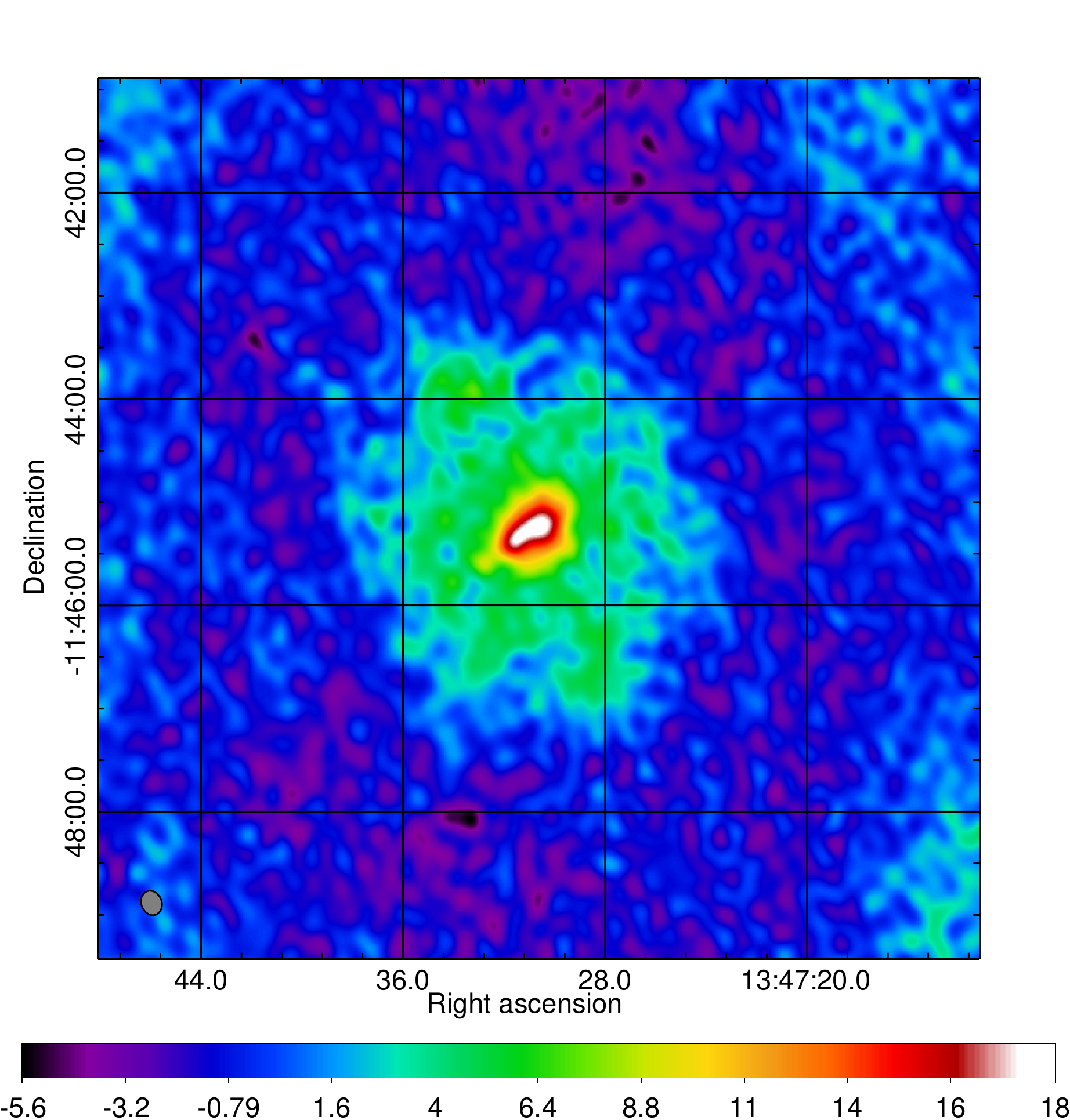}
\end{center}
\caption{Dirty map of RX~J1347.5$-$1145 using ``robust'' visibility
weighting.  The synthesized beam is
shown in grey in the lower left.  The map is in units of Compton 
$y \times 10^{13}/$beam.}
\label{fig:dirty}
\end{figure*}

We next apply a phase shift to the CARMA-15 visibilities to establish
a single phase center for all ten data sets.  At this stage, 
all of the visibilities can be combined to form a 
single dirty map (Figure~\ref{fig:dirty}).  However, this map is
difficult to interpret because it includes data
with dramatically different primary beams, and because noise at
small angular scales masks the arcminute-scale emission.  
In order to proceed, we must use some image deconvolution method such as the
Maximum Entropy Method (MEM) or CLEAN.  For simplicity, we choose to
build a model of the cluster iteratively using the CLEAN algorithm,
though we note that MEM also holds promise for future work.

\begin{figure*}
\begin{center}
\includegraphics[height=4in]{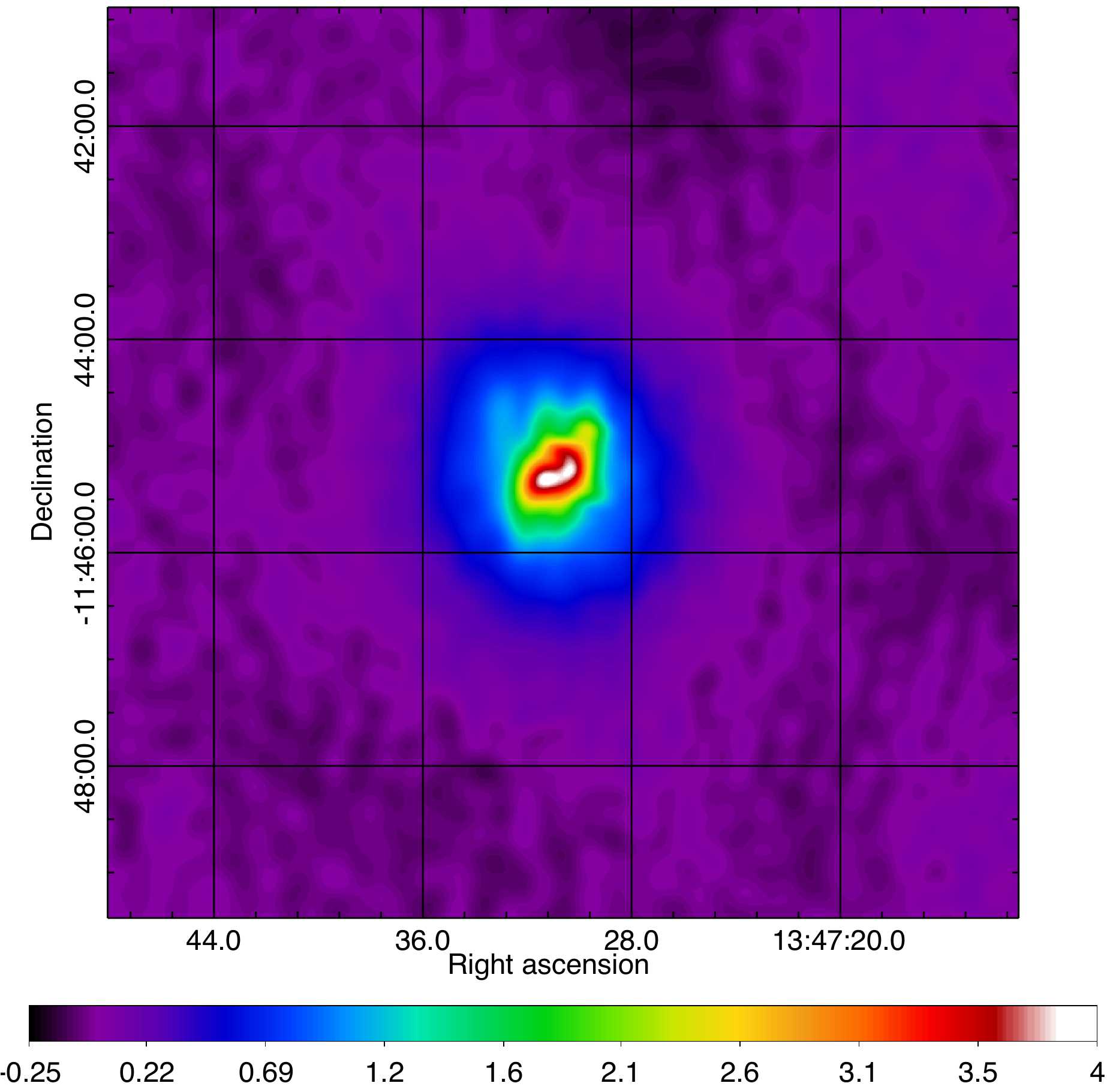}
\end{center}
\caption{CLEAN map of RX~J1347.5$-$1145.  The map is in units of Compton
$y \times 10^{15}$/pixel, where the pixel size is 
$0.5^{\prime\prime}\times 0.5^{\prime\prime}$.  The smallest CLEAN beam
used to construct the model is $10^{\prime\prime}.6 \times 16^{\prime\prime}.9$}.
\label{fig:clean}
\end{figure*}

Our algorithm successively builds up a single model of the cluster
in image space by CLEANing individual data sets.  We begin by making a dirty map 
from the first data set $y_1(u,v)$.  We then run the H\"{o}gbom CLEAN 
algorithm \citep{Hogbom74} on this map with gain 0.05, stopping at 
$2.5\sigma$.  The CLEAN components are corrected for the appropriate
primary beam, and a restored map is generated.  This restored 
map (in units of Compton $y$ per pixel) is our initial cluster model.
The cluster model is then multiplied by the primary beam of the next data 
set, $y_2(u,v)$, and the result is Fourier transformed and subtracted 
from $y_2(u,v)$.  From these model-subtracted visibilities, a dirty map 
is produced and H\"{o}gbom CLEANed.  The CLEAN components are corrected 
for the primary beam and restored, and the restored map is added to
the cluster model.  We repeat this process on all ten data sets, resulting
in a cluster model that incorporates information from all of our data.
We continue refining the model by making additional passes through all ten
data sets until each model-subtracted dirty map is consistent with noise.  
Three iterations are found to be sufficient.  Our final cluster model is
then added to the weighted average of the ten model-subtracted dirty maps
to form a CLEAN map (Figure~\ref{fig:clean}).
Though the resolution of the map is not well-defined, the smallest 
restoring beam of the ten used to construct the model 
is 10\farcs6 $\times$ 16\farcs9.

%The ten single-array dirty maps and residuals
%are shown in Figure~\ref{fig:resid} \red{Include this figure?}. 
%\begin{figure*}
%\begin{center}
%\includegraphics[width=6.5in]{pm5sig.pdf}\\
%\includegraphics[width=6.5in]{pm5sig_resid.pdf}
%\end{center}
%\caption{Dirty maps (top two rows) and residual maps (bottom two rows) for each 
%of the ten CARMA data sets.  Both the dirty maps and residual maps are ordered from 
%top left to bottom right by increasing synthesized beam solid angle, and
%the primary beam half power contours for each map are outlined in black.  
%The color scale ranges from $-5\sigma$ to $+5\sigma$ for each map, such that
%several of the dirty maps are saturated.
%The residuals are consistent with noise, indicating that the
%iterative CLEAN technique has converged on a self-consistent model.}
%\label{fig:resid}
%\end{figure*}

\begin{figure*}
\begin{center}
\includegraphics[height=4in]{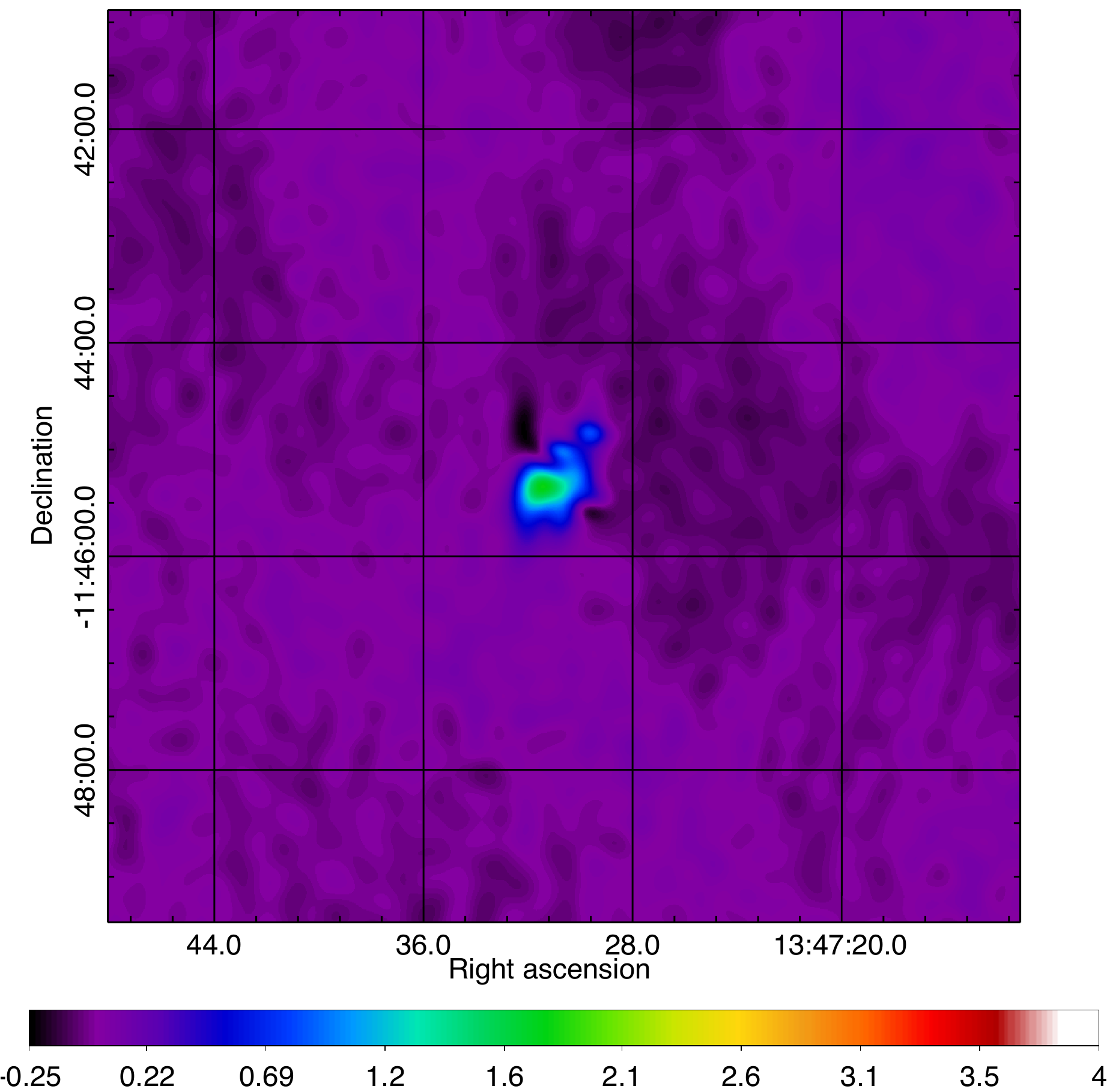}
\end{center}
\caption{CLEAN map of RX~J1347.5$-$1145 with the scaled relaxed X-ray pressure
profile subtracted from the visibilities.  The map is in units of Compton
$y \times 10^{15}$/pixel, where the pixel size is 
$0.5^{\prime\prime}\times 0.5^{\prime\prime}$.}
\label{fig:disturbed}
\end{figure*}

To assess the accuracy and flux recovery of our deconvolution technique, 
we construct a heuristic model of RX~J1347.5$-$1145.  We generate mock
data using this model, repeat the
procedure described above on the simulated visibilities, and compare the
results to the input model.
Modeling the SZ signal requires knowledge of the density $n_{\mathrm{e}}$ and 
the temperature $T_{\mathrm{e}}$ of the ICM, both of which can be approximated
using the results of \citet{Allen02} (hereafter \citetalias{Allen02}) derived 
from \emph{Chandra} X-ray data. 
For the electron density, we use the \citetalias{Allen02} density fit 
within the central region with an $\alpha=-2.331$ power law in the outer 
region inferred from the \citetalias{Allen02} surface brightness profile fit. 
We piece together the inner and outer regions by requiring that $n_{\mathrm{e}}(r)$ be 
continuous.  For the electron temperature, we use the 
\citetalias{Allen02} measured temperature profile out to the outer radius of 
their largest bin, and use the average temperature of 
$12.0$~keV at larger radius.  We assign a temperature of $18.0$~keV 
in the shocked region of the southeast quadrant.  Finally,
we truncate both the temperature and density profiles at $r_{200}$.

Using these approximations for the electron density and temperature, we
integrate along the line of sight to produce a simulated Compton $y$ 
map, and Fourier transform to produce simulated visibilities $y_{\mathrm{sim}}(u,v)$.  
We then randomize the phases in our visibility data $y(u,v)$ to produce
simulated noise $y_{\mathrm{noise}}(u,v)$ and run our iterative CLEAN algorithm on
$y_{\mathrm{sim}}(u,v) + y_{\mathrm{noise}}(u,v)$.  We find that our algorithm accurately
reproduces the morphology of the input model, and that it recovers
69\%, 78\%, and 92\% of the integrated flux within $r_{200}$, $r_{500}$, and $r_{2500}$, where the
$r_\Delta$ values are determined from the \citet{Allen02} NFW model fit.  The
flux recovery ratio at larger radii is highly sensitive to the assumed electron 
density power law index.  This is due to the fact that $r_{200}$ for this 
cluster corresponds to an angular scale of 5\farcm75,
comparable to the 10\farcm7 FWHM 3.5~m primary beam at 1~cm.

\section{Results}
\label{sec:results}

\begin{figure*}
\begin{center}
\includegraphics*[width=0.9\textwidth]{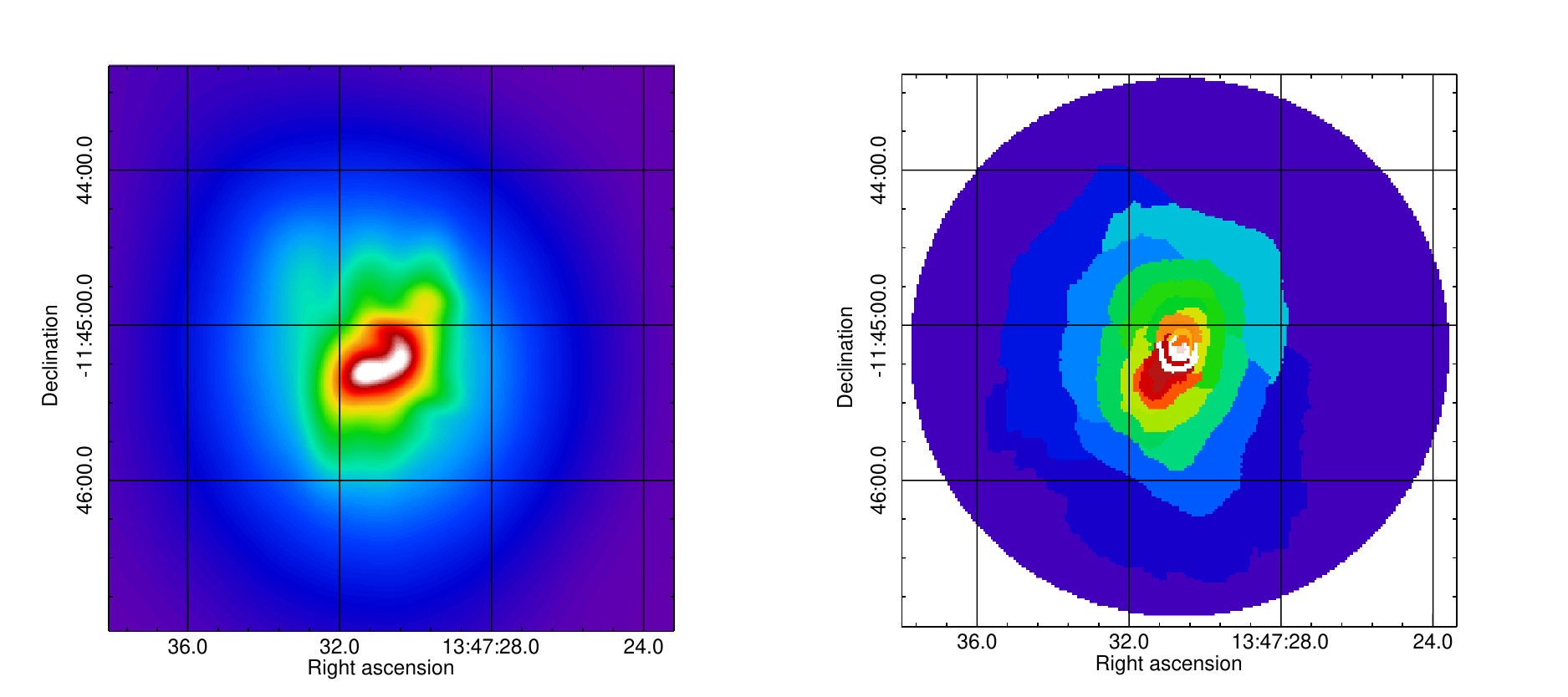}
\caption{
The CLEANed CARMA SZ map (left) cropped and scaled for direct
comparison with the published X-ray-derived integrated pressure
map from \citet{Bradac08} (right). 
}
\label{fig:multiband}
\end{center}
\end{figure*}

The map shown in Figure~\ref{fig:clean} can be understood as a
relaxed cluster SZ signal with additional sub-arcminute structure in the 
southeast quadrant imparted by the merger event.  
The imaging of both the extended and compact structure in our SZ map 
is of significantly higher fidelity than previous measurements
due to the ability of CARMA to remove the central point source
and to the large angular dynamic range of the combined arrays.
In contrast with previous work \citep{Mason10,Komatsu00}, we
find that the peaks of the SZ and X-ray signals are coincident.
Figure~\ref{fig:multiband} shows a comparison between the
CARMA SZ map and the \chandra\ X-ray pressure map \citep{Bradac08}.
Though a full multi-wavelength reconstruction is beyond the scope
of this paper, it is obvious that the two techniques produce
maps with consistent morphologies.  

To separate the relaxed and disturbed components,
we make use of the pressure profile fit to the \chandra\ X-ray data
described in \citet{Allen08}, in which the southeast quadrant was
excised.  We first project the pressure profile
along the line of sight to produce an integrated pressure map.  For each sub-array, 
we multiply this map by the appropriate primary beam, convert to Compton $y$, 
Fourier transform, and subtract the result from the visibility data.  We then
iteratively build a model of the remaining SZ signal, following the procedure
described in Section~\ref{sec:decon}.  We allow the X-ray pressure profile
to be scaled by a multiplicative constant to compensate for cluster projection effects
and calibration errors.  The scale factor is chosen so as to produce
no net CARMA-8 (arcminute-scale) signal in the iteratively-determined model.

Removing this estimate of the relaxed signal from
our visibility data allows us to focus on the sub-arcminute-scale
signal resulting from the merger event.  The result is shown in 
Figure~\ref{fig:disturbed}.  The total Compton $y$ signal recovered in this
map, which corresponds to the fraction of the thermal energy in the cluster
ICM associated with the merger-related substructure, is $9.1$\% of 
the total recovered from the map in Figure~\ref{fig:clean}.

The value of the multiplicitive constant by which the X-ray pressure
is scaled can also yield information about the characteristics of the cluster.  
Since the scale factor is determined by requiring consistency with the
CARMA-8 data, its value depends upon the three-dimensional morphology
of the ICM on large angular scales.
Due to the different dependencies of the X-ray surface brightness and the
SZ signal on density, a cluster more (less) extended
along the line of sight than in the plane of the sky would yield a scale
factor of greater than (less than) one \citep[see e.g.,][]{Grego04}.
For this cluster, we find a scale factor of 0.58, indicating a
cluster which is strongly compressed along the line-of-sight direction.
A similar ratio is reported in \citet{Bonamente12} using the CARMA-8
data reported here.  \citet{Chakra08}, using a combination of X-ray and SZ data, 
also find that RX~J1347.5$-$1145 is compressed along the line of sight
with an axis ratio of $\sim 5$.  
A simple comparison of the X-ray surface brightness profile and our data,
assuming an isothermal cluster, implies compression along the line of sight by
a factor of roughly three-to-one---a fairly extreme value, though less than
suggested by \citet{Chakra08}.  

Clumping of the ICM, i.e., a systematic discrepancy between 
$\langle n_{\mathrm{e}}^2(r) \rangle$ and $\langle n_{\mathrm{e}}(r) \rangle^2$, 
could also lead to differences between the 
SZ and X-ray signals. Clumping is observed in the outskirts of
simulated clusters \citep{Nagai11}, and is implied by observations
of flattened entropy profiles and gas mass fractions apparently in excess of
the cosmic mean in some clusters \citep[e.g.,][]{Simionescu11}.  However,
in both cases the clumping occurs at large cluster radii (at least $>r_{500}$),
and is thus unlikely to affect the normalization of the X-ray model used
here (which was fit to data at $r<r_{500}$) at the level required to 
explain the offset.  

The difference between the CARMA and \chandra\ pressure
estimates could also arise in principle from calibration or
systematic errors in the SZ or X-ray data.  As a cross-check of our CARMA
calibration, we compared our data to previous BIMA SZ measurements
reported in \citet{Bonamente08}, finding consistent integrated $Y$
values and binned visibility data.  Since the \chandra\ calibration 
is unlikely to be mistaken at this level, and since
major mergers such as RX~J1347.5$-$1145 are not expected to be spherically
symmetric, we suggest that line-of-sight compression is most likely to be 
the dominant effect.  However, ICM clumping and calibration errors may
also be contributing to the discrepancy; a more complete explanation will require
a joint analysis of the two data sets.

%Using data from XMM-Newton, \citet{Gitti04} measure X-ray spectroscopic 
%temperatures which are significantly smaller than those estimated by \chandra\ 
%\citep{Allen08,Bradac08}.  A concurrent fit to the \chandra\ surface brightness
%data and the \citet{Gitti04} temperature profile produces a pressure profile
%that is consistent with our SZ map (i.e.\ the best-fit scale factor is
%$\sim 1.0$).  However, the combination of XMM-Newton's extended point spread 
%function and RX~J1347.5-1145's X-ray-luminous cool core suggest that the
%XMM-Newton temperature measurement is likely an underestimate.

\begin{table*}
\centering
\caption{CARMA Observations of RX~J1347.5$-$1145.}
\begin{tabular}{|c|c|c|c|c|c|c|c|c|}
\hline
\hline
Array & Ant 1 & Ant 2 & Frequency & $uv$ cutoff & Noise & Minor axis & Major axis & Beam P.A. \\
      &       &       & (GHz)     & (k$\lambda$)& mJy/beam & arcsec & arcsec & Degrees \\
\hline
CARMA-8  & 3.5m & 3.5m & 31        & 2000        & 0.16  & 100.1      & 128.8 & -33.4 \\
CARMA-15 & 10.4m  & 10.4m  & 90        & 10000       & 0.38  & 13.4       & 16.8  &  72.9 \\
CARMA-15 & 10.4m  & 6.1m & 90        & 10000       & 0.19  & 12.8       & 16.5  & -44.8 \\
CARMA-15 & 6.1m & 6.1m & 90        & 10000       & 0.30  & 15.1       & 37.4  &   3.0 \\
CARMA-23 & 10.4m  & 10.4m  & 86        & 10000       & 0.22  & 13.9       & 14.6  & -80.2 \\
CARMA-23 & 10.4m  & 6.1m & 86        & 10000       & 0.14  & 14.4       & 16.4  & -65.3 \\
CARMA-23 & 6.1m & 6.1m & 86        & 10000       & 0.26  & 14.9       & 37.3  &   4.4 \\
CARMA-23 & 10.4m  & 3.5m & 88        & 10000       & 0.80  & 16.2       & 18.3  &  59.1 \\
CARMA-23 & 6.1m & 3.5m & 88        & 10000       & 0.76  & 10.6       & 16.9  & -81.1 \\
CARMA-23 & 3.5m & 3.5m & 88        & 5000        & 2.64  & 31.4       & 67.5  & -44.3 \\
\hline
\hline
\end{tabular}
\label{tab:obs}
\end{table*}

\section{Conclusions and Future Work}
\label{sec:conclusions}

We have demonstrated the ability of CARMA to measure the SZ signature of
galaxy clusters at high sensitivity across a wide range of angular scales.  
By combining data from three CARMA configurations and two frequency
bands, we measure the arcminute-scale SZ signal as well as the pressure 
substructure of RX~J1347.5$-$1145.  The large angular dynamic range of
the CARMA data and the capability to constrain and remove point sources
make our measurement a significant improvement in image fidelity over 
previous work \citep[e.g.,][]{Korngut11}.
By comparing our data to the X-ray measurements of \citet{Allen08}, we are
able to determine that $\sim 9$\% of the SZ signal is localized in the
disturbed region of the cluster, and that the system is likely compressed
along the line of sight relative to the plane of the sky.

Although our results demonstrate that CARMA in its current form is a
highly capable SZ instrument, several upgrades will soon bring about
significant enhancements.  An effort is currently underway to equip all
antennas with 1~cm receivers and expand the correlator bandwidth to
8~GHz for a 23-element array.  The 3~mm CARMA-23 array described in this work
provided sensitivity from $uv$ radii of $\sim 2.0$ to $10$k$\lambda$;
the upgraded array placed in the same configuration but operated at 1~cm
will provide higher-sensitivity coverage from $\sim 0.35$ to $3.3$k$\lambda$.
As shown in Figure~\ref{fig:radbin}, this corresponds to the portion of
the $uv$ plane where the SZ signal is large.  A planned upgrade to 
more sensitive 3~mm receivers will allow the SZ signal at finer 
angular scales to be measured more precisely.  

The sensitivity to smaller angular scale SZ structures provided CARMA's  
larger telescopes can be directed toward regions of interest, as was 
done in this study by pointing them toward the region of hot gas to 
the southeast of the cluster center.  With the increased sensitivity 
enabled by the ongoing upgrades, this technique can be used to search 
for shock-enhanced features in the outskirts of clusters.  Mosaicking 
can also be used to provide sensitivity to small-scale structures 
over the entire cluster.  

Taken together, these upgrades and observing strategies will
allow CARMA to image clusters precisely and efficiently
over a wide angular dynamic range, making it possible to fully exploit
the power of the SZ effect as a probe of cluster astrophysics and 
precision cosmology.

\section*{Acknowledgements}

We thank Evan Million for providing the X-ray pressure data from \citet{Bradac08}.  
We also thank Maru{\v s}a Brada{\v c} and Myriam Gitti for useful discussions.

Support for CARMA construction was derived from the Gordon and Betty Moore
Foundation, the Kenneth T. and Eileen L. Norris Foundation, the James
S. McDonnell Foundation, the Associates of the California Institute of
Technology, the University of Chicago, the states of California,
Illinois, and Maryland, and the National Science Foundation. Ongoing
CARMA development and operations are supported by the National Science
Foundation under a cooperative agreement, including grant AST-0838187
at the University of Chicago, and by the CARMA partner universities.
Partial support is provided by NSF Physics Frontier Center grant
PHY-1125897 to the Kavli Institute of Cosmological Physics.  D.~P.~M.~was 
supported for part of this work by NASA through Hubble Fellowship
grant HST-HF-51259.01.  

Finally, we thank the CARMA staff for making 
the 23-element commissioning observations possible.

{\it Facilities:} \facility{CARMA}

\bibliography{sz.bib}
\bibliographystyle{hapj}

\end{document}